\documentclass{iopart}
\usepackage{epsfig}
\usepackage{amsfonts}
\usepackage{amssymb}
\usepackage{mathptmx}
\usepackage{eucal}

\newcommand{\sgn}{\,\mathrm{sgn}\,}
\newcommand{\re}{\,\mathrm{Re}\,}
\newcommand{\im}{\,\mathrm{Im}\,}
\begin{document}

\title{Multiparticle tunnelling in diffusive superconducting junctions}

\author{E V Bezuglyi${\dag\ddag}$, A S Vasenko${\dag\S}$, E N Bratus'$\ddag$,
V S Shumeiko$\dag$ and G Wendin$\dag$}
\address{$\dag$ Chalmers University of Technology, S-41296 G\"oteborg, Sweden}

\address{$\ddag$ B Verkin Institute for Low Temperature Physics and Engineering,
Kharkov 61103, Ukraine}

\address{$\S$ Department of Physics, Moscow State University, Moscow
119992, Russia}

\ead{eugene.bezuglyi@gmail.com}

\begin{abstract}
We formulate a theoretical framework to describe multiparticle current
transport in planar superconducting tunnel junctions with diffusive
electrodes. The approach is based on direct solving of quasiclassical
Keldysh-Green function equations for nonequilibrium superconductors, and
consists of a combination of a circuit theory analysis and improved
perturbation expansion. The theory predicts much greater scaling parameter
for the subharmo\-nic gap structure of the tunnel current in diffusive
junctions compared to the one in ballistic junctions and mesoscopic
constrictions with the same barrier transparency.
\end{abstract}


\maketitle

\section{Introduction}

Multiparticle tunnelling (MPT) is known to be a mechanism of dissipative
current transport in superconducting tunnel junctions at the subgap applied
voltage $eV<2\Delta$, and at small temperature $T\ll \Delta$ \cite{MPT}. It
has been shown in \cite{Bratus95,Cuevas96} that the MPT is completely
equivalent to the {\em coherent} multiple Andreev reflection mechanism (MAR)
\cite{Averin95,Arnold87}. Here we use the term MPT to emphasize the low
transparency, tunnel junction limit, leaving the term MAR for a general case
of transparent weak links. Each MPT event can be considered as a chain of
multiple Andreev reflections \cite{KBT} accompanied by the transfer of $n$
electrons through the tunnel barrier and eventually results in the creation
of two quasiparticles which contribute to the dissipative current. This
process manifests itself in the set of the current steps at $eV = 2\Delta/n$
-- the subharmonic gap structure (SGS) -- which is commonly observed in
planar junctions with tunnel barriers
\cite{Burstein,Marcus,Cristiano,Gubrun2001}, and in tunable mesoscopic
constrictions in the tunnelling regime \cite{Post94,Scheer97,Jan2000}. The
theory predicts the scaling $\approx D/2$ between the neighboring current
steps in the tunnelling SGS, where $D$ is the bare transparency of the
junction tunnel barrier \cite{MPT,Bratus95,Cuevas96,Averin95}. However, in
the experiment, the SGS scaling parameter is generally much larger. A common
explanation for this enhancement is the imperfection of the junction
insulating layer \cite{KBT,Marcus,Kle}. In our previous paper
\cite{BezuglyRC06} the attention was drawn to an alternative explanation:
effect of disorder in planar junction electrodes; it was predicted  that the
SGS scaling parameter for planar diffusive junctions is enhanced by a factor
$\sim \xi_0/\ell$, or even $\sim \xi_0^2/\ell d$ depending on the junction
geometry ($\xi_0$ is the coherence length, $\ell$ is the elastic mean free
path, $d$ is the thickness of electrodes).

In this paper we present a detailed theory of the MPT in planar Josephson
tunnel junctions with diffusive thin-film electrodes and extensively discuss
the details of the behaviour of the MPT currents and the relevant
asymptotical methods. The theory is based on the direct solving of the
diffusive equations of nonequilibrium superconductivity \cite{LOnoneq}. The
main difficulty with this approach is related to essentially nonstationary
character of the Josephson tunnel transport. While analytical and numerical
methods are well developed for solving stationary Usadel equation
\cite{Belzig} and nonequilibrium Keldysh-Usadel equation \cite{Bezugly2000},
the nonstationary problem was so far studied only numerically
\cite{Cuevas06}. The central model assumption made in this paper and relevant
for the tunnel regime concerns discrimination of nonzero harmonics of the
Green's functions and the distribution function. This approximation turns the
originally difficult problem into an analytically tractable one, and at the
same time it captures all qualitative, and to large extent quantitative
properties of the tunnelling SGS. Within this approximation we are able to
develop a relatively simple and physically appealing calculation scheme,
which combines the improved iterative procedure for evaluating the tunnelling
density of states (DOS), and the circuit theory methods
\cite{Bezugly2000,Nazarov99} for evaluating the dc current.

\begin{figure}[tb]
\begin{center}
\epsfxsize=6.5cm\epsffile{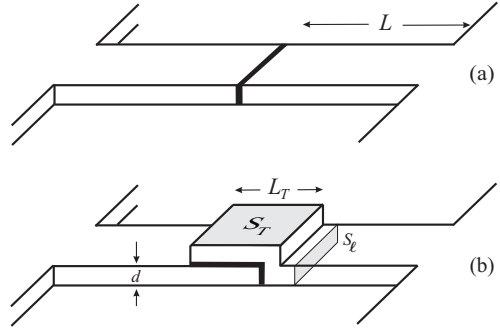}
\end{center} \vspace{-6mm}
\caption{One-dimensional (a) and planar (b) models of the tunnel junction.}
\label{model}
\end{figure}

The resulting physical picture of the MPT is as follows: The tunnelling
processes create the local nonzero DOS inside the bulk energy gap in the
vicinity of the tunnel junction. This allows quasiparticles to overcome the
energy gap at a small applied voltage in several steps, by repeated bouncing
between the junction electrodes (MAR). The spectral current through the
energy gap, which determines the net charge current, is calculated by
considering an effective circuit theory network representing the tunnelling
process.

The effect of substantial enhancement of the SGS scaling factor can be
qualitatively understood from a mesoscopic picture of the diffusive
tunnelling transport, namely, tunnelling through a set of independent
quasi-ballistic conducting channels with randomly distributed transparencies
\cite{Nazarov99,Beenakker}. Within this picture, the contribution of each
channel can be evaluated using the ballistic MAR theory
\cite{Bratus95,Cuevas96,Averin95}. In constrictions with length $L \gg \ell$
the transparencies are spread over the interval $\thicksim (L/\ell)D \gg D$
\cite{Beenakker94}, which implies that the junction transparency, and hence
the SGS scaling factor, are effectively enhanced by the factor $L/\ell$. This
explanation, however, is valid only for short constrictions, $L\ll\xi_0$,
while it does not apply to planar tunnel junctions with overlapping thin-film
electrodes commonly used in experiments and shown in figure \ref{model}. In
these structures, massive pads (reservoirs) are fairly far from the junction
($L \gg \xi_0$); in such a situation the effective junction length is defined
by the scale of spatial variation of the Green's function, i.e., by the
coherence length. If the size $L_{\rm T}$ of the overlapping parts of the
electrodes is comparable to the electrode thickness $d$, then the junction
can be considered as an effectively one-dimensional (1D) one (see figure
\ref{model}(a)); in this case, the SGS enhancement factor becomes
$\xi_0/\ell$. If the junction cross-section is much larger than the
cross-section of the electrodes (see figure \ref{model}(b)), the current
concentrates not at the junction but rather in electrodes \cite{VolkovSIN},
and an additional enhancement factor $\xi_0/d$ appears in the SGS scaling,
which coincides with the result of rigorous calculation in this paper.
Remarkably, this enhancement concerns only the multiparticle currents, while
the single-particle current is not affected and is proportional to the bare
transparency $D$ \cite{Werthamer}.

The paper is organized as follows. In its major part we develop a theory for
the 1D junctions, figure \ref{model}(a), and discuss the extension to the
planar junctions, figure \ref{model}(b), towards the end, in section
\ref{Secplanar}. We start with discussion of basic equations and adopted
approximations in section \ref{Sec1D}. Then we construct the circuit theory
in section \ref{Seccircuit}, and develop a perturbation theory for the DOS in
section \ref{SecDOS}. The MPT currents are calculated in sections
\ref{Secthresholds} and \ref{Secpeaks}; the latter section includes also the
calculation of the excess current. The effect of neglected harmonics in the
Keldysh and Green's functions is evaluated in section \ref{Secharmonics}. In
section \ref{Diss} we discuss the results and possible implications of the
theory.

\section{1D junction model}\label{Sec1D}

\subsection{Basic equations}

The model of tunnel junction we are first going to study is depicted in
figure \ref{model}(a) and consists of a tunnel barrier with the transparency
$D$ attached to bulk superconducting electrodes via two superconducting leads
($-L < x <0$ and $0<x<L$). We will consider a diffusive limit, in which the
elastic scattering length $\ell$ is much smaller than the coherence length
$\xi_0 = \sqrt{\mathcal{D}/2\Delta}$, where $\mathcal D$ is the diffusion
coefficient (we assume $\hbar = k_{\rm B} = 1$). We assume the length $L$ of
the leads to be much larger than $\xi_0$, and their width to be much smaller
than the Josephson penetration depth which implies homogeneity of the current
along the junction. Similar model has been considered in \cite{BBG} in study
of the dc Josephson effect in tunnel structures.

Under these conditions, the microscopic calculation of the electric current
$I(t)$ requires solution of the diffusive equations of nonequilibrium
superconductivity \cite{LOnoneq} for the $4 \times 4$ matrix two-time
Keldysh-Green's function $\check{G}(x,t_1,t_2)$ in the leads,
\begin{eqnarray}
\left[\check{H}, \circ\check{G}\right] = \rmi \mathcal{D}
\partial_x\check{J}, \qquad \check{J} = \check{G}\circ\partial_x
\check{G}, \qquad \check{G}\circ\check{G} = \delta(t_1 - t_2),\label{GK}
\\ \nonumber
\check{H} = \bigl[\rmi\sigma_z \partial_{t_1} - e\varphi +
\hat{\Delta}(t_1)\bigr]\delta(t_1-t_2), \qquad \hat{\Delta} =
\rme^{\rmi\sigma_z\phi}\rmi\sigma_y\Delta,
\end{eqnarray}
where $\varphi$ is the electric potential, $\Delta$ and $\phi$ are the
modulus and the phase of the order parameter, respectively, $\sigma_i$ are
the Pauli matrices, $\partial_x$ denotes partial derivative over the variable
$x$ and
\begin{equation}
\check{G} = \left(\begin{array}{ccc} \hat{g}^{\rm R} & \hat{G}^{\rm K} \\
0 & \hat{g}^{\rm A}
\end{array}\right), \qquad \hat{G}^{\rm K} = \hat{g}^{\rm R} \circ\hat{f}
- \hat{f}\circ \hat{g}^{\rm A}. \label{defG}
\end{equation}
Here $\hat{g}^{\rm R,A}$ are the $2 \times 2$ Nambu matrix retarded and
advanced Green's functions, and $\hat{f} = f + \sigma_z f_-$ is the matrix
distribution function (we use `check' for $4 \times 4$ and `hat' for $2
\times 2$ matrices). The multiplication procedure in \eref{GK} and
\eref{defG} involves the time convolution
\begin{equation}
(A\circ B)(t_1, t_2) = \int^{+\infty}_{-\infty} A(t_1, t)B(t, t_2) \rmd
t.\nonumber
\end{equation}

For arbitrary tunnel barrier, the function $\check{G}$ and the matrix current
$\check{J}$ at the left ($x=-0$) and the right ($x=+0$) sides of the tunnel
junction are connected via the generalized boundary condition by Nazarov
\cite{Nazarov99},
\begin{equation}
\check{J}_{-0} = \check{J}_{+0} = \frac{1}{2g_{\rm N} R} \int_0^1
\frac{D\rho(D)\, \rmd D \bigl[\check{G}_{-0}, \circ
\check{G}_{+0}\bigr]}{1+\frac{D}{4}\bigl(\bigl\{\check{G}_{-0},
\circ \check{G}_{+0}\bigr\}-2\bigr)},\label{Naz}
\end{equation}
where $\rho(D)$ is the distribution of the transparencies of the conducting
channels of the barrier ($\int_0^1 D\rho(D)\, \rmd D =1$), $R$ is the
junction resistance and $g_{\rm N}$ is the normal conductivity of the leads
per unit length. Assuming the absence of high-transparent channels with $D
\sim 1$ and considering $\rho(D)$ to be localized around the small value of
$D \ll 1$ (tunnel limit), we can neglect the anti-commutator term in
\eref{Naz}, thus arriving to the Kupriyanov-Lukichev's boundary condition
\cite{KL},
\begin{equation}
\check{J}_{-0} = \check{J}_{+0} = (2g_{\rm N} R)^{-1} \bigl[\check{G}_{-0},
\circ \check{G}_{+0}\bigr].\label{KL}
\end{equation}
The electric current is related to the Keldysh component of the matrix
current $\check{J}$ as $I(t)= (\pi g_{\rm N}/4e) \Tr \sigma_z \hat{J}^{\rm
K}(x,t,t)$, and thus it can be expressed through the boundary value
$\check{J}_0$,
\begin{equation}
I(t)=({\pi }/{8eR}) \Tr \sigma_z \bigl[\check{G}_{-0},\circ
{\check{G}_{+0}}\bigr]^{\rm K}(t,t). \label{Curr2}
\end{equation}

Equations (\ref{GK}) can be decomposed into the diffusion equations
for the Green's functions,
\begin{equation}
\bigl[\hat{H}, \circ \hat{g}\bigr] = \rmi \mathcal{D}\partial_x \hat{J},
\qquad \hat{J} = \hat{g}\circ\partial_x \hat{g}, \qquad \hat{g}\circ\hat{g} =
\delta(t_1 - t_2),\label{Usadel}
\end{equation}
and the equation for the Keldysh component $\hat{G}^{\rm K}$,
\numparts
\begin{eqnarray}
\bigl[\hat{H}, \circ \hat{G}^{\rm K}\bigr] = \rmi \mathcal{D}\partial_x
\hat{J}^{\rm K}, \qquad \hat{J}^{\rm K}  = \hat{g}^{\rm R}
\circ\partial_x\hat{G}^{\rm K} + \hat{G}^{\rm K} \circ\partial_x \hat{g}^{\rm
A}, \label{KineticG1}
\\
\hat{g}^{\rm R} \circ \hat{G}^{\rm K} + \hat{G}^{\rm K} \circ \hat{g}^{\rm A}
= 0. \label{G_Knorm1}
\end{eqnarray}
\endnumparts
The boundary conditions for the functions $\hat{g}$ and $\hat{G}^{\rm K}$ at
the tunnel barrier follow from \eref{KL},
\numparts
\begin{eqnarray}
\hat{J}_{0}  = (W / \xi_0)\bigl[\hat{g}_{-0}, \circ {\hat{g}}_{+0}\bigr],
\label{BoundaryG}
\\
\hat{J}_{0}^{\rm K}  = (W / \xi_0)\bigl[\check{G}_{-0}, \circ
{\check{G}}_{+0}\bigr]^{\rm K}.
 \label{BoundaryGK}
\end{eqnarray}
\endnumparts

In \eref{BoundaryG} and \eref{BoundaryGK}, the transparency parameter $W$ is
defined as
\begin{equation} \label{W}
W = {R(\xi_0)}/{2R} = ({3\xi_0}/{4\ell}) D \gg D,
\end{equation}
where $R(\xi_0) = \xi_0/g_{\rm N}$ is the normal resistance of the piece of
the lead with the length $\xi_0$. It has been shown in \cite{BBG} that it is
the parameter $W$ rather than the barrier transparency $D$ that plays the
role of a true transparency parameter in diffusive tunnel junctions. We will
consider the limit $W \ll 1$, which corresponds to the conventional
tunnelling concept. In this case, according to \eref{BoundaryG} and
\eref{BoundaryGK}, the gradients of all functions are small. Within the
tunnel model, which assumes $W$ to be the smallest parameter in the theory,
these gradients are neglected, and the functions $\hat{g}$ and $\hat{f}$ are
taken local-equilibrium within the leads. In our consideration, we will lift
this assumption and suppose the local-equilibrium form of these functions
only within the bulk electrodes (reservoirs). Attributing the reference point
for the phase, $\phi=0$, to the left electrode, $x=-L$, these functions in
the right electrode, $x=L$, are given by relations
\numparts
\begin{eqnarray}
\hat{g}(E,t)  =  \sigma_z u(E+\sigma_z eV) + \rmi
\rme^{\rmi\sigma_z\phi(t)}\sigma_y v(E) , \label{gS}
\\
(u,v) ={(E,\Delta)}/\xi,\qquad \xi^{\rm R,A}=[(E\pm \rmi 0)^2 -
\Delta^2]^{1/2}, \label{xi}
\\
\hat{f}(E) = \tanh[(E+\sigma_z eV)/2T], \label{f0}
\end{eqnarray}
\endnumparts
written in terms of the mixed Wigner representation $A(E,t)$ of the
two-time functions,
\begin{eqnarray} 
A(t_1, t_2) = \int_{-\infty}^{+\infty} \frac{\rmd E}{2 \pi }\, \rme^{-\rmi E
(t_1 - t_2)} A(E, t),\nonumber
\end{eqnarray}
where the variable $E$ has the meaning of the quasiparticle energy, and $t =
(t_1 + t_2)/2$ is a real time. Similar equations, with $\phi=0$ and $V=0$,
apply to the left electrode, $x=-L$.

Because of the small value of the tunnelling parameter $W$ one can neglect
variations of the superconducting phase along the leads, as well as the
charge imbalance function $f_-$ proportional to a small electric field
penetrating the superconducting leads. Furthermore, the small value of the
superfluid momentum in the superconducting leads, $p_s \sim W$ \cite{BBG},
allows us to neglect a small effect of the energy gap suppression by the
superfluid momentum ($\sim p_s^{4/3} \sim W^{4/3}$ \cite{LO1}). Within such
an approximation, the coefficients in \eref{GK} at the left lead, $x<0$, are
time-independent functions. At $x>0$, applying the gauge transformation
\cite{Artemenko} $\widetilde{\check{G}}(t_1,t_2) = S^\dagger(t_1)
\check{G}(t_1,t_2) S(t_2)$ with a unitary operator $S(t) =\exp[\rmi
\sigma_z\phi(t)/2]$ to the function $\check{G}$, we exclude the
time-dependent phase and the electric potential from the equations for the
function $\widetilde{\check{G}}$ and corresponding boundary conditions at
$x=L$, which then become similar to the equations for $\check{G}(x)$ at $x<0$
and the boundary conditions at $x=-L$. This results in the symmetry relation
$\widetilde{\check{G}}(x) = \check{G}(-x)$, which allows us to replace the
function $\check{G}_{+0}$ in the boundary condition \eref{KL} and in the
expression \eref{Curr2} for the electric current by the inverse gauge
transformation of the function $ \check{G}_{-0}$, $\check{G}_{+0} \to
\overline{\check{G}}_{-0} \equiv   S(t_1) \widetilde{\check{G}}_{+0}
S^\dagger(t_2) = S(t_1) \check{G}_{-0} S^\dagger(t_2)$.

As the result, the problem is reduced to the solution of a static equation
within the left lead for the function $\check{G}(x,t_1,t_2)$, completed with
the time-dependent boundary condition \eref{KL} at the tunnel barrier.
Similar approach is used in the theory of ballistic point contacts
\cite{Bratus95} where the Josephson coupling is described by an effective
time-dependent matching condition for the gauge-transformed Bogolyubov-de
Gennes equations in the leads.

It is convenient to expand all functions over harmonics of the Josephson
frequency, $A(E,t)=\sum_m A(E,m)\exp(-2\rmi eV mt)$, using the following
rules for representation of the pro\-ducts and gauge-transformed values,
\begin{eqnarray}
(A \circ B)(E,m) \label{harm}
\\ \nonumber
=\sum\nolimits_{m'} A\left[E + eV (m - m'), m'\right] B\left(E - eV m', m -
m'\right),
\\
\overline{\hat{A}}(E,m)= \left(\begin{array}{ccc} A_{11}(E + eV,m), &
A_{12}(E,m + 1)
\\
A_{21}(E,m - 1),& A_{22}(E-eV,m) \end{array}\right). \label{harmgauge}
\end{eqnarray}
In such a representation, the expression \eref{Curr2} for the dc current $I$
has the following form
\begin{equation} \label{I0}
\eqalign{ I =  \int_{-\infty }^\infty \frac{\rmd E}{16eR} \Tr ( \hat{h} \circ
\overline{\hat{G}^{\rm K}} - \overline{\hat{h}} \circ \hat{G}^{\rm K})(E,0)
\\
=\int_{-\infty }^\infty \frac{\rmd E}{16eR} \Tr
\sum\nolimits_m\bigl[\hat{h}(E,m)\overline{\hat{G}^{\rm K}}(E,-m)
\\
-\overline{\hat{h}}(E,m) \hat{G}^{\rm K}(E,-m)\bigr], \qquad \hat{h} =
\sigma_z \hat{g}^{\rm R} - \hat{g}^{\rm A} \sigma_z, }
\end{equation}
where all functions are taken at the boundary $x=-0$. We will adopt the same
convention in the most of following equations and assume the spatial
coordinate to be taken at the boundary.

\subsection{Zero-harmonic model}

Solving a system of nonlinear differential equations
\eref{Usadel}--\eref{BoundaryGK} generally can be fulfilled only numerically
even in the 1D case. The analytical solutions can be constructed in the
adiabatic limit of small applied voltage $eV\ll\Delta$ \cite{Bezuglui2005}.
To make the problem tractable at larger voltages $eV\sim\Delta$, we make use
of the observation that the amplitudes of high-order harmonics of the
function $\check{G}$ are small in the tunnelling limit $W\ll 1$: the
amplitude of the $m$th harmonic decreases with $m$ as $W^m$. This suggests
that zero harmonics $m=0$ play the key role in \eref{I0}, while the
high-order harmonics are neglected. Thus we adopt an approximation scheme, in
which only the zero harmonics of the functions $\hat{g}$ and $\hat{G}^{\rm
K}$ are kept. It turns out that such an approximation is sufficiently
powerful to recover all specific features of the MPT currents, and to give
satisfactory description of the SGS of the net tunnel current. Furthermore,
our analysis of the correction due to the first harmonics in section
\ref{Secharmonics} shows that the zero-harmonic model may give rather good
quantitative agreement with the result of full numerical calculation.

For the matrix structure of the zero harmonic of the function $\hat{g}$ in
\eref{I0}, we adopt the form
\begin{equation}
\hat{g}(E,0) = \sigma_z u(E) + \rmi\sigma_y v(E), \qquad u^2 - v^2 =1,
\label{g0}
\end{equation}
which is similar to the Green's function structure \eref{gS} in the left
electrode, though $u$ and $v$ differ from their equilibrium values in
\eref{xi}. It is possible to prove, using the normalization condition in
\eref{Usadel}, that the zero harmonic of matrix $\hat{g}$ is traceless, and
its $\sigma_x$-component is much smaller (at least by $W^2$) than the
``main'' components $u$ and $v$, which have zero order in the parameter $W$.
Within the same approximation, the Keldysh function has the form
$\hat{G}^{\rm K}(E,0) = 2f(\sigma_z N+\rmi\sigma_y M)$ , where $N(E) = \re
u^{\rm R}$ is the density of states (DOS) normalized to its value in the
normal state and $M(E)= \re v^{\rm R}$. In what follows, we will express the
advanced functions through the retarded ones, $(u,v)^{\rm A} = - (u,v)^{\rm
R\ast}$, using the relation $\hat{g}^{\rm A} = -\sigma_z \hat{g}^{\rm
R\dagger}\sigma_z$, and omit the superscript $\rm R$, assuming all Green's
functions to be retarded.

Retaining only zero harmonics of the functions $\hat{g}$ and $\hat{G}^{\rm
K}$ in \eref{I0}, we find that only the diagonal parts ${h}(E)$ and ${G}^{\rm
K}(E)$ of the corresponding matrices enter the dc current
\begin{eqnarray}\label{Iderivation}
I &= \int_{-\infty}^{\infty}\frac{\rmd E}{32 eR} \Tr\sum_{k=\pm1} (1 + k
\sigma_z)[{h}(E) {G}^{\rm K}(E + k eV)
\\
&- {h}(E + k eV){G}^{\rm K}(E)].\nonumber
\end{eqnarray}
By introducing the distribution function $n = \case12(1-f)$ which approaches
the Fermi function $n_{\rm F}$ in equilibrium, equation \eref{Iderivation}
exactly transforms to the standard form for the tunnel current,
\begin{equation}\label{I00}
I = \int_{-\infty }^\infty \frac{\rmd E}{eR} N(E) N(E-eV) [n(E-eV)-n(E)],
\end{equation}
with that essential difference that the DOS and the distribution function are
not given, but are to be computed from the Keldysh-Green's function
equations. To zero order in the tunnelling parameter, the DOS has the BCS
form $N_{\rm S}(E) = \re (E/\xi)$, and the distribution function is the
equilibrium one, $n=n_{\rm F}$. In this approximation, equation \eref{I00}
recovers the single-particle current of the tunnel model \cite{Werthamer}. At
zero temperature this current acquires the form
\begin{equation}
I = \int_\Delta^{eV-\Delta} \frac{\rmd E}{eR}\, N_{\rm S}(E)N_{\rm S}(E -eV)
\label{I001}
\end{equation}
and turns to zero at $eV<2\Delta$, having a sharp onset at $eV =
2\Delta$, $I_1(2\Delta) = \pi\Delta/2eR$.

To calculate the current at smaller, subgap voltages $eV<2\Delta$, one has to
calculate the tunnelling corrections to the BCS DOS, and to find the
nonequilibrium distribution function.

\section{Circuit representation of boundary condition}\label{Seccircuit}

We start with evaluation of the distribution function and develop a circuit
theory approach to derive a general analytical equation for the current
\eref{I00} assuming the DOS to be modified in a close vicinity of the tunnel
barrier.

\subsection{Kinetic equation and boundary condition}

Using zero harmonic of \eref{KineticG1} and \eref{BoundaryGK}, we obtain the
diffusive kinetic equation and the boundary condition,
\begin{eqnarray}
\partial_x\left(D_+\partial_x n\right) = 0,\label{kinetic1}
\\
D_+\partial_x n\bigr|_{x=0} = (2W/\xi_0) \sum\nolimits_{k=\pm 1}NN_k
\left(n_k - n\right)\bigr|_{x=0},\label{kinetic2}
\end{eqnarray}
where $D_+ = \case12(1 + |u|^2 - |v|^2)$ is the dimensionless diffusion
coefficient, and the subscript $k$ denotes the energy shift: $n_k(E) \equiv
n(E_k) = n(E + keV)$. It follows from \eref{kinetic1} that $D_+\partial_x n =
\rm{const}$; this constant value can be found from \eref{kinetic2}. At $x\gg
\xi _{0}$ the coefficient $D_+$ approaches the BCS form $D_{\rm S} =
\Theta(|E| - \Delta)$ ($\Theta(x)$ is the Heaviside step function) and
exactly turns to $D_{\rm S}$ at the reservoir, $x=-L$. This implies that at
subgap energies $|E|<\Delta$, the quasiparticle probability current
$D_+\partial_x n$ turns to zero along the whole lead, and the distribution
function is spatially homogeneous. Physically, this manifests complete
Andreev reflection in terms of quasiparticle flows in diffusive structures.

Outside the gap, equation \eref{kinetic1} has no bound solutions: the
distribution function grows linearly with $x$ far from the junction. Such a
growth is limited in practice by inelastic colli\-sions, which provide
relaxation of $n(x,E)$ to the equilibrium value $n_{\rm F}$ at the distance
of inelastic scattering length $l_{\varepsilon}$. To simplify the problem, we
consider, instead of including a complicated inelastic collision term, a
junction geometry with short enough leads having the length $L\ll
l_{\varepsilon}$ (but still $L \gg \xi_0$) and connected at $x=\pm L$ to
equilibrium reservoirs. Within this model, the Green's functions, which
change at $x\sim \xi _{0}$, are not affected by the finite length of the
leads, while the reservoirs impose the equilibrium boundary conditions for
the distribution function, $n(\pm L) = n_{\rm F}$. At the same time we can
neglect inelastic collisions inside the leads. Obviously, this model
describes a qualitative pattern of inelastic relaxation in very long channel
($L \gg \ell_\varepsilon$) with substitution $L \to \ell_{\varepsilon }$ in
our results.

Generally, spatial variation of $n(x)$ at $|E|>\Delta$ has two scales: linear
$x$-dependence at $x \sim L$ and fast but small variations near the junction
due to spatial dependence of the Green's function. Neglecting these
variations within the main approximation in the parameters $W$ and $\xi_0/L$,
we arrive at the relation $D_+\partial_x n = [{n(0) - n_{\rm F}}]L^{-1}$.
Substituting it into the boundary condition \eref{kinetic2} and accounting
for $D_+\partial_x n = 0$ at $|E| < \Delta$, we obtain the equation
\begin{equation}
\Theta(|E| - \Delta)(n - n_{\rm F}) = r \sum\nolimits_{k=\pm 1}NN_k\left(n_k
- n\right).\label{circuit_th}
\end{equation}
In this and following equations, the functions are taken at the boundary $x =
- 0$. Equation \eref{circuit_th} represents a recurrence relation between the
values of $n(E)$ at the energies shifted by $eV$; the nonequilibrium
parameter $r$ is defined as
\begin{equation}
r = 2LW/\xi_0 = R_{\rm N}/R, \label{beta}
\end{equation}
where $R_{\rm N} = L/g_{\rm N}$ is the normal resistance of one lead. In
practice, the tunnel resistance much exceeds $R_{\rm N}$, and the
nonequilibrium parameter is small, $r \ll 1$, which implies that at the
energies $|E| > \Delta$, the distribution function is always close to the
Fermi function.

\subsection{Circuit theory}

We split the integral in \eref{I00} into pieces of length $eV$,
\begin{eqnarray}
I =  \int_{-\infty}^\infty \frac{\rmd E}{e R} j_0(E) = \int_0^{eV}\frac{\rmd
E}{e R}  J(E),\label{I002}
\\
\nonumber J = \sum\nolimits_{k=-\infty}^{\infty} j_k, \quad j_k = \left(
n_{k-1} - n_k \right) \rho_k^{-1}, \quad \rho_k^{-1} = N_k N_{k-1}.
\end{eqnarray}
In these notations the recurrence relation \eref{circuit_th} reads
\begin{equation}
\Theta(|E_k| - \Delta) \left[n_k - n_{\rm F}(E_k)\right] = r (j_k -
j_{k+1}).\label{recurr}
\end{equation}
\begin{figure}[tb]
\begin{center}
\epsfxsize=8cm\epsffile{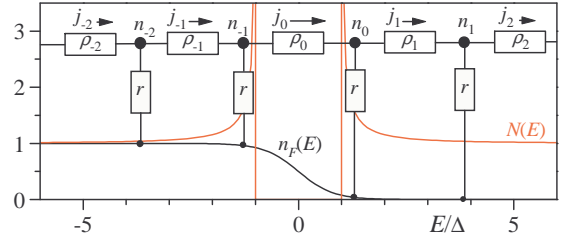} \end{center} \vspace{-5mm}
\caption{Circuit theory network with the period $eV$ representing charge
transport in diffusive tunnel junctions, $eV = 2.5\Delta$.} \label{circuit}
\end{figure}

A convenient interpretation of equations \eref{I002} and \eref{recurr} in
terms of the circuit theory \cite{Bezugly2000} is given by an infinite
network in the energy space with the period $eV$, graphically presented in
figure \ref{circuit}. The electric current spectral density\ $J(E)$ consists
of partial currents $j_k$ which flow through the chain of tunnel
``resistors'' $\rho_k$ connected to adjacent nodes of the network having
``potentials'' $n_k$ and $n_{k-1}$. At $|E|>\Delta$, the nodes are also
attached to the distributed ``equilibrium source'' $n_{\rm F}(E)$ through
equal resistors $r$. In this representation the recurrence relation
\eref{recurr} has the meaning of ``Kirchhoff rules'' for partial currents.

Below we assume the equilibrium quasiparticle distribution $n(E)= n_{\rm
F}(E)$ at $|E|>\Delta$, neglecting  effect of small resistors $r$. In this
limit, the currents $j_k$ outside the energy gap, $|E_k| > \Delta$, vanish at
zero temperature since $n_{\rm F}$ is piecewise constant. At $T \neq 0$,
these currents describe the effect of thermal excitations. The subgap
spectral current is conserved, $j_k = j_\Delta ={\rm const}, \;\; k = 1 -
N_-, \ldots, N_+$ (as a consequence of \eref{recurr}), and can be easily
computed,
\begin{eqnarray}\nonumber
j_\Delta = \left[n_{\rm F}(E_{-N_-}) - n_{\rm F}(E_{N_+})\right]
\rho_\Delta^{-1},
\\
N_\pm(E) = {\rm Int}\,\left[(\Delta \mp E)/eV\right]+1,\qquad \rho_\Delta(E)
= \sum\nolimits_{k= 1-N_- }^{N_+} \rho_k,\nonumber
\end{eqnarray}
where the integers $\pm N_\pm$ are the numbers of the nodes
outside the gap nearest to the gap edges, Int$(x)$ denotes integer
part of $x$, and the quantity $\rho_\Delta(E)$ has the meaning of
the net subgap resistance. The subgap distribution function reads
\begin{equation} \label{nnet}
n(E)=n_{\rm F}(E_{N_+})+ \left[n_{\rm F}(E_{-N_-}) - n_{\rm
F}(E_{N_+})\right]\sum\nolimits_{k=1}^{N_+}\rho_k\rho_\Delta^{-1}.
\end{equation}

The resulting electric current can be now written in a general form,
\begin{equation}\label{Inet}
I = \int_0^{eV}\frac{\rmd E}{eR}(N_- + N_+)j_\Delta
+2\int_\Delta^\infty\frac{\rmd E}{eR}\frac{n_{\rm F}(E) - n_{\rm
F}(E_1)}{\rho_1},
\end{equation}
valid for arbitrary voltages and temperatures. Here the first term describes
the subgap current, and the second term -- the current of thermal
excitations.

The magnitude of the subgap current is fully determined by the net subgap
resistance;  the current is blocked when this resistance is infinite,
$\rho_\Delta = \infty$, which  happens when DOS turns to zero. According to
\eref{Inet}, the amount $N_- + N_+$ of the resistors contributing to the
subgap resistance gives the amount of electric charge (in units of $e$)
transferred during the tunnelling event. Thus, the circuit with one subgap
resistor represents the single-particle tunnelling, which can exist only at
$eV>2\Delta$; in this case, equation \eref{Inet} reduces to \eref{I001}. At
$eV<2\Delta$, the subgap circuit should consist of at least two resistors
(two-particle tunnelling). However, for the BCS DOS this current is blocked,
and to evaluate the current one has to calculate the tunnelling correction to
the DOS within the gap by solving the Green's function equations.

\section{Planar junctions}\label{Secplanar}

In this section we will discuss the extension of our approach to a more
practical case of planar tunnel junction sketched in figure \ref{model}(b).
This 2D case is more complex; however, it is possible to reduce this problem
to the 1D case by formulating effective boundary conditions at the junction
following the method suggested by Volkov \cite{VolkovSIN}.

Let us suppose that the size of the junction $L_{\rm T}$ exceeds the
coherence length, $L_{\rm T} \gg \xi_0$, and, simultaneously, does not exceed
the Josephson penetration depth. Then the function $\check{G}$ in the
left-hand side (lhs) of the kinetic equation $ [\check{H}, \circ \check{G}] =
\rmi \mathcal{D}\nabla \check{J}$ is approximately constant within the
junction banks (parts of the junction leads of lengths $L_{\rm T}$ beneath
and above the insulator). Then, integrating this equation over the volume of
the bottom bank, using the boundary conditions \eref{KL} and denoting the
cross-section area of the lead as $S_\ell$, the lead thickness as $d$, and
the area of the junction as $S_{\rm T}$, we obtain
\begin{eqnarray}
S_T d [\check{H}, \circ \check{G}] = \rmi \mathcal{D}\bigl\{ S_{\rm T}
(W/\xi_0)[\check{G}_{-0}, \circ \check{G}_{+0}] - S_\ell \check{J}_\ell
\bigr\}, \;\; {\rm{or}}  \nonumber
\\
\left[\check{H}, \circ \check{G}\right] = 2\rmi\Delta \bigl\{ \widetilde{W}
[\check{G}_{-0}, \circ \check{G}_{+0}] - (\xi_0^2/L_{\rm T}) \check{J}_\ell
\bigr\}, \label{Bound1}
\end{eqnarray}
where $\pm 0$ denotes top and bottom side of the barrier, $\check{J}_\ell$ is
the value of the matrix current at the lead cross-section adjoining the
junction, and the tunnelling parameter $\widetilde{W}$ is defined as
\begin{equation}\label{Wtilde}
\widetilde{W} = W(\xi_0/d)= ({3\xi_0^2}/{4\ell d}) D.
\end{equation}
As soon as $\xi_0 \ll L_{\rm T}$ and $\xi_0 \check{J}_\ell \sim W$, the last
term in \eref{Bound1} can be assumed to be the smallest one and thus
neglected. However, this is only true for the Green's component of
\eref{Bound1} \cite{VolkovSIN},
\begin{eqnarray}
[\check{H}, \circ \check{g}] = 2\rmi\Delta \widetilde{W} [\check{g}_{-0},
\circ \check{g}_{+0}], \label{Boundg}
\end{eqnarray}
whereas for the Keldysh component, the diagonal part of the lhs of
\eref{Bound1} turns to zero (we consider only zero harmonics), and therefore
the boundary condition for the diagonal part of $\hat{J}_\ell^{\rm K}$, which
is proportional to $D_+\partial_x f$, has the form
\begin{eqnarray}
\hat{J}_\ell^{\rm K} = (W_{\rm f}/\xi_0)[\check{G}_{-0}, \circ
\check{G}_{+0}]^{\rm K} , \qquad W_{\rm f} = W(L_{\rm T}/d) \gg
\widetilde{W}. \label{Boundf}
\end{eqnarray}
Equation \eref{Boundf} is the boundary condition for the distribution
function, which is to be used as described in previous section. Solving the
kinetic equation within the lead and assuming $L \gg L_{\rm T}$, we arrive at
the equation similar to \eref{recurr} with the same parameter of
nonequilibrium,
\begin{eqnarray} 
r = \frac{2W_{\rm f} L}{\xi_0} = 2L\frac{R(\xi_0)}{2\xi_0 R} \frac{L_{\rm
T}}{d} = L \frac{\rho}{S_{\rm T} R}\frac{L_{\rm T}}{d}= \frac{L\rho}{S_\ell
R} = \frac{R_{\rm N}}{ R},\nonumber
\end{eqnarray}
where $\rho$ is the specific conductivity of the leads.

\section{Perturbation theory for the Green's functions}\label{SecDOS}

To calculate the DOS within the next approximation with respect to the
parameter $W$, we solve the Usadel equation for the Green's function
$\hat{g}$ following from \eref{Usadel}. Introducing usual parametrization
$\hat{g}=\sigma_z\exp (\sigma_x\theta)$ and the dimensionless coordinate $z$,
we arrive at the equation for the spectral angle $\theta$,
\begin{eqnarray}
\sinh[\theta(z) - \theta_{\rm S}] = \rmi \theta''(z)\,\sinh\theta_{\rm S} ,
\qquad z = x/\xi_0 \label{Eqtheta}
\end{eqnarray}
(the prime sign denotes the derivative over $z$). With exponential accuracy,
the solution of \eref{Eqtheta} at $z<0$ can be approximated by the formula
for a semi-infinite superconducting wire \cite{BV}
\begin{eqnarray}
\tanh[({\theta(z) - \theta_{\rm S}})/{4}] = \tanh[({\theta(-0) - \theta_{\rm
S} })/{4}] \exp(kz),  \label{Soltheta}
\end{eqnarray}
where $k^{-1}(E) ={\sqrt{\rmi\sinh\theta_{\rm S}}}$. Equation \eref{Soltheta}
describes the decay of perturbations of the spectral functions at distances
$\gtrsim \xi_0$ from the barrier, where the spectral angle approaches its
bulk value $\theta_{\rm S} ={\rm arctanh} (\Delta/E)$. The boundary condition
for the spectral angle follows from \eref{BoundaryG},
\begin{eqnarray}
\theta' +W\sinh\theta (\cosh\theta_1 +
\cosh\theta_{-1})\bigr|_{z=-0}=0. \label{Boundtheta}
\end{eqnarray}
Then the boundary value of $\theta$ can be found from the finite-difference
equation following from \eref{Boundtheta} and \eref{Soltheta},
\begin{equation}
2k\sinh [({\theta_{\rm S} - \theta})/{2}] =W\sinh\theta (\cosh\theta_1 +
\cosh\theta_{-1}). \label{Eqtheta1D}
\end{equation}
A similar result for the spectral angle in planar junction banks follows from
\eref{Boundg},
\begin{eqnarray}
k^2 \sinh(\theta_{\rm S}-\theta) = \widetilde{W} \sinh\theta (\cosh\theta_1 +
\cosh\theta_{-1}). \label{EqthetaPl}
\end{eqnarray}

In what follows, we will simultaneously discuss both of the junction
geometries, using common notation $W$ for both transparency parameters $W$
and $\widetilde{W}$ and assuming this quantity to be defined by \eref{W} or
\eref{Wtilde} depending on context.

\begin{figure}[tb]
\begin{center}
\epsfxsize=8cm\epsffile{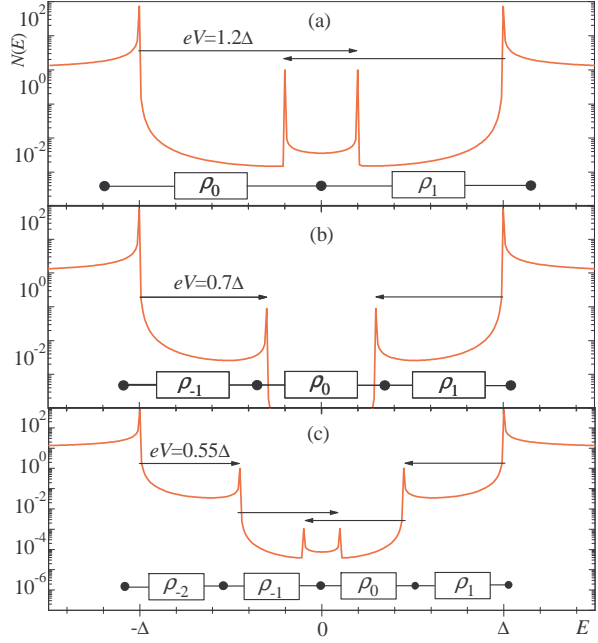} \end{center} \vspace{-5mm}
\caption{Numerically computed DOS and typical subgap circuits describing the
two-particle current at $eV=1.2\Delta$ (a), the three-particle current at $eV
= 0.7\Delta$ (b), and the four-particle current at $eV=0.55 \Delta$ (c), for
the tunnelling parameter $W =10^{-3}$.} \label{DOS}
\end{figure}

\subsection{Simple perturbation theory}\label{SecSPT}

Due to the presence of the small parameter $W$ in the right-hand side (rhs)
of \eref{Eqtheta1D}, one can suggest the following perturbation correction
for $\theta$ to first order in $W$,
\begin{equation}
\theta =\theta_{\rm S} - Wk^{-1} \sinh\theta_{\rm S} \left(\cosh \theta_{\rm
S,1} + \cosh \theta_{\rm S,-1} \right), \label{SPT}
\end{equation}
and similar for \eref{EqthetaPl}. This results in the following expression
for the DOS within the BCS gap,
\begin{eqnarray}
N(E)= \re (\cosh \theta)= W\bigl|\Delta/\xi\bigr|^{a}(N_{\rm S,1}+N_{\rm
S,-1}), \label{DOSSPT}
\\ \nonumber |\xi| =\sqrt{\Delta^2-E^2}, \qquad
a =\cases{ 5/2, & \textrm{1D junction;}
\\ 3, & \textrm{planar junction.}}
\end{eqnarray}
Such approximation will be referred to as the result of a simple perturbation
theory (SPT).

As follows from \eref{DOSSPT}, the tunnelling coupling extends the DOS inside
the gap over the distance $eV$ from the gap edges, and scales it down by the
factor $W$, as shown in figure \ref{DOS}. It is clear from \eref{DOSSPT} that
with decreasing voltage, at $eV<\Delta$, the gap will open in the DOS (see
figure \ref{DOS}(b)), and further iteration of the finite-difference
equations \eref{Eqtheta1D} or \eref{EqthetaPl} for the spectral angle is
required. As the result of this iterative scheme, the DOS at small enough
$eV$ acquires a staircase structure in the energy space: two ladders
descending from the bulk gap edges inside the subgap region; the height of
$n$th step is $W^n$, and the width is $eV$. In the middle of the gap, $|E| <
\Delta - (n-1)eV$ (assuming $eV < 2\Delta/n$), there is a plateau with the
height $\approx 2W^n$ (see figures \ref{DOS}(a) and \ref{DOS}(c)). While $eV$
decreases, the plateau expands until its size becomes equal to $2eV$, then a
new pair of steps emerges, which happens when $n$ is even. We recall that
this deformation of the DOS occurs only locally, at the distance $x\sim
\xi_0$ from the junction.

\subsection{Improved perturbation theory}\label{SecIPT}

The subgap DOS in \eref{DOSSPT} possesses singularities at the gap edges $|E|
= \Delta$, which causes a divergence of the subgap current at relatively
large voltage, as we will see later. To eliminate the divergence, we need to
apply an improved perturbation theory (IPT) to \eref{Eqtheta1D} and
\eref{EqthetaPl}, in which nonlinearity of the recurrence relations is fully
taken into account. First, we consider an approximation to \eref{Eqtheta1D},
in which the non-singular terms $\cosh\theta_{\pm 1}$ are replaced with their
BCS values, but we do not suppose that the difference $\theta - \theta_{\rm
S}$ is small,
\begin{eqnarray} \label{IPT_I2}
&\sinh \left[({\theta_{\rm S} - \theta})/{2} \right] =Wk^{-1}g(E, eV)
\sinh\theta ,
\\
&g(E, eV) = \case12\left(\cosh\theta_{\rm S,1} + \cosh\theta_{\rm
S,-1}\right).\nonumber
\end{eqnarray}

In the vicinity of the point $E = \Delta$ the function $g(E, eV)$ is regular
and thus can be approximated by $g(\Delta, eV)$ (it is sufficient to consider
only positive values of $E$ due to the symmetry of the spectral functions).
Within the region $|E - \Delta| \ll \Delta$, the spectral angles $\theta$ and
$\theta_{\rm S}$ are large, therefore we hold only large exponents
$\exp\theta$ and $\exp\theta_{\rm S}$ in the hyperbolic functions in the rhs
of \eref{IPT_I2}, and use the asymptotic expression $\exp\theta_{\rm S}
\approx [2\Delta/(E - \Delta)]^{1/2}$. Then, introducing a dimensionless
energy variable $\epsilon$ and a normalized spectral function $z(\epsilon)$,
\begin{equation}\label{IPT_2not}
\eqalign{ \epsilon = ({\Delta -E})/{2\Delta p^2} , \qquad p(eV) =[\case12 W^2
g^2(eV)]^{1/3}, \label{p_Notations}
\\
z(\epsilon)=\rmi p\exp\theta , \qquad \exp \theta_{\rm S} = (\rmi
p\sqrt\epsilon)^{-1} }
\end{equation}
we reduce \eref{IPT_I2} to a numerical algebraic equation
\begin{equation}\label{IPT_2zeq}
z^3 + \left(z\sqrt{\epsilon} - 1\right)^2 = 0.
\end{equation}
The relevant solution $z(\epsilon)$ of \eref{IPT_2zeq} is determined by the
requirement for the asymptotic beha\-viour at $\epsilon \gg 1$ to coincide
with the energy dependence $z(\epsilon) \approx\epsilon^{-1/2} +
\rmi\epsilon^{-5/4}$ given by the direct perturbative expansion \eref{SPT}.
For planar geometry, we directly obtain the function $z(\epsilon)$ and the
scaling parameter $p$ from \eref{EqthetaPl},
\begin{equation} \label{p_planar}
z(\epsilon)=(\epsilon-\rmi)^{-1/2}, \qquad p(eV)=[Wg(eV)]^{1/2}.
\end{equation}
The resulting DOS in the region $|E - \Delta| \ll \Delta$,
\begin{equation}\label{IPT_2DOS}
N(E, eV) = [{2p(eV)}]^{-1} \im \,z(\epsilon),
\end{equation}
approaches a finite value $N(\Delta, eV) \sim W^{-b}(eV -
2\Delta)^{-b/2}$ at $E = \Delta$, where
\begin{equation} \label{b}
b=\cases{ 2/3, & \textrm{1D junction;}
\\ 1/2, & \textrm{planar junction.}}
\end{equation}
However, in the vicinity of the specific voltage value $eV = 2\Delta$ the DOS
diverges, and the calculation procedure must be further improved. The problem
is caused by the fact that at $eV = 2\Delta$ both the energies $E$ and $E
-eV$ in \eref{Eqtheta1D} or \eref{EqthetaPl} approach the gap edges $\Delta$
and $-\Delta$, respectively. Therefore one must solve the equation not only
for $N(E)$, but for $N(E - eV)$ as well. To this end we consider the
recurrence \eref{Eqtheta1D} for these two energies and replace the
nonsingular terms $\cosh\theta_1$ and $\cosh\theta_{-2}$ by their BCS values,
\begin{equation}\label{EqIPT1}
\eqalign {\sinh \frac{\theta_{\rm S} - \theta}{2} = \frac{W\sinh\theta}{2k}
(\cosh\theta_{-1} + \cosh\theta_{\rm S,1}),  \\  \sinh \frac{\theta_{S,-1} -
\theta_{-1}}{2} = \frac{W\sinh\theta_{-1}}{2k_{-1}} (\cosh\theta_{\rm
S,-2}+\cosh\theta).}
\end{equation}
Using again the fact that the spectral angles are large by modulus in the
region $|E - \Delta| \ll \Delta$, we hold only large exponents $\exp\theta$,
$\exp\theta_{\rm S} \approx [{2\Delta/(E - \Delta)}]^{1/2}$ and
$\exp(-\theta_{-1})$, $\exp(-\theta_{S,-1}) \approx [2\Delta/(eV -
\Delta-E)]^{1/2}$ in the rhs of \eref{EqIPT1}. Then, introducing
dimensionless energy and voltage variables $\epsilon$ and $\Omega$, and
normalized spectral functions $z(\epsilon)$ and $\overline{z}(\epsilon)$,
\begin{equation}\label{IPT_Notations}
\eqalign{ \epsilon = ({E - \Delta})/{2\Delta q^2}, \qquad  \Omega =({eV -
2\Delta})/{ 2\Delta q^2},
\\
\exp\theta = zq^{-1}, \qquad \exp(-\theta_{-1}) = \overline{z} q^{-1}, \qquad
q = W^{2/5}/2.}
\end{equation}
we obtain algebraic equations for the functions $z(\epsilon)$ and
$\overline{z}(\epsilon)$, which reduce to a single equation for the function
$z(\epsilon)$,
\begin{eqnarray} \label{EqIPT2}
\bigl[z^3(1-zR)^{-2} +\rmi\overline{R}^2\bigr]^2 +\rmi z \bigl[(1-zR)^{2}-
4z\overline{R} \bigr]=0,
\end{eqnarray}
where $R(\epsilon) = \sqrt{\epsilon + \rmi 0}$ and $\overline{R}(\epsilon) =
R(\Omega - \epsilon)$. The function $\overline{z}$ can be then found as
$\overline{z} = (1 - zR)/\sqrt{\rmi z^3}$. For planar geometry, we obtain the
equations
\begin{eqnarray}\label{IPTpl}
(1-z^2 R^2)^2 (z + \rmi\overline{R}^2) + \rmi z^4 = 0, \qquad
\overline{z}=({1-z^2 R^2})/{\rmi z^2},
\end{eqnarray}
and the scaling parameter $q=({W}/{4})^{1/3}$. According to the definition of
$z(\epsilon)$ in \eref{IPT_Notations}, the solutions of \eref{EqIPT2} and
\eref{IPTpl} are related to the boundary values of the DOS as
\begin{eqnarray} \label{DOS_IPT}
N(E) &= (2q)^{-1}\re \,z,\qquad N_{-1}(E)= (2q)^{-1} \re \,\overline{z}.
\end{eqnarray}
The results of computation of the DOS based on the numerical solution of the
recurrences (\ref{Eqtheta1D}) and (\ref{EqthetaPl}) and shown in figure
\ref{DOS}, quantitatively confirm the results of our asymptotic analysis.

\section{Multiparticle currents: thresholds}\label{Secthresholds}

\subsection{Two-particle current}

The existence of the subgap states enables quasiparticles to overcome the
energy gap at $eV < 2\Delta$ via two steps involving intermediate Andreev
reflection at the energy $|E|<\Delta$. The population $n(E)$ of the
intermediate state is generally non-equilibrium, because the subgap
quasiparticles cannot access the equilibrium electrodes. In terms of the
circuit approach, the node $k=0$ is disconnected from the equilibrium source,
and the subgap current flows through two resistances $\rho_0$ and $\rho_1$
(two-particle current), see figure \ref{DOS}(a). The corresponding partial
currents are equal, $j_0 = j_1 = [{n_{\rm F}(E_1)-n_{\rm F}(E_{-1})}]
/({\rho_0 + \rho_1})$, and their contribution to $I(V)$ is confined to the
energy region $0< E < eV-\Delta$ (a similar contribution at $\Delta < E < eV$
comes from $j_0$ and $j_{-1}$), which leads to the following expression for
the two-particle current,
\begin{eqnarray}\nonumber
I_2 = \frac{4}{eR}\int_0^{eV-\Delta} \frac{\rmd E} {\rho_0 + \rho_1}, \qquad
\Delta \leq eV < 2\Delta.
\end{eqnarray}
Within the SPT approximation for the subgap DOS function $N$, this
is equivalent to equation
\begin{equation}\label{deltaI3}
I_2 = \frac{4W}{eR} \int_0^{eV-\Delta} \rmd E
\left|\frac{\Delta}{\xi}\right|^{a}{\frac{|E_1 E_{-1}|}{\xi_1 \xi_{-1}}},
\end{equation}
where $\xi(E) = [(E+\rmi 0)^2-\Delta^2]^{1/2}$ according to \eref{xi}. The
current $I_2$ increases with voltage and diverges at $eV = 2\Delta$ which is
the result of the mentioned DOS singularity at $E=\Delta$.

At $eV = \Delta$, the two-particle current possesses a threshold. In the
vicinity of the threshold, $eV = \Delta + \Omega$, $0 < \Omega \ll \Delta$,
equation \eref{deltaI3} simplifies, giving the current threshold value
\begin{eqnarray} \label{I2thr}
&I_2(\Delta) = \frac{2W\Delta}{eR}\int_0^{\Omega} \frac{\rmd
E}{\sqrt{\Omega^2 - E^2}}= \frac{\pi W\Delta}{eR} = 2WI_1(2\Delta).
\end{eqnarray}

\subsection{Three-particle current}

At $eV < \Delta$, a minigap opens in the DOS around the zero energy (see
figure \ref{DOS}(b)), however, since the number of subgap resistors increases
up to three: $\rho_{-1}$, $\rho_0$ and $\rho_1$ (three-particle current), the
current across the minigap will persist as long as the network period exceeds
the minigap size, $eV> 2(\Delta-eV)$, i.e., at $eV > 2\Delta/3$. The
corresponding partial currents are equal, $j_{-1} = j_0 = j_1 = [{n_{\rm
F}(E_1)-n_{\rm F}(E_{-2})}]/({\rho_{-1} + \rho_0 + \rho_1})$, and their
contribution to $I(V)$ is confined to the energy region $\Delta - eV < E <
2eV-\Delta$, which leads to the following expression for the three-particle
current at zero temperature,
\begin{equation}\label{I3>}
I_3 = \frac{3}{eR}\int_{\Delta - eV}^{2eV-\Delta} \frac{\rmd E} {\rho_{-1} +
\rho_0 + \rho_1}, \qquad 2\Delta/3 \leq eV < \Delta.
\end{equation}
Taking the subgap DOS functions $N$ and $N_{-1}$ in the SPT approximation
\eref{DOSSPT}, we see that the central resistance $\rho_0 \sim W^{-2}$ is the
largest. Retaining only this resistance, we get
\begin{eqnarray}\label{I3>>}
I_3 &= \frac{3 W^2}{eR} \int_{\Delta - eV}^{2eV-\Delta} \rmd
E\left|\frac{\Delta^2}{\xi \xi_{-1}}\right|^{a} {\frac{|E_1 E_{-2}|}{\xi_1
\xi_{-2}}}.
\end{eqnarray}
While approaching the voltage value $eV=\Delta$, the current $I_3$ infinitely
increases due to decreasing distance between the upper integration limit and
the singular point $E=\Delta$. Calculating \eref{I3>>} in the vicinity of the
threshold $eV = 2\Delta/3$, we obtain
\begin{eqnarray} \label{I3thr}
I_3\left({2\Delta}/{3}\right) = \frac{3\pi\Delta W^2}{2eR} \left(
\frac{9}{8} \right)^{a} \approx 2W I_2(\Delta).
\end{eqnarray}

\subsection{Four-particle current}

At $eV<2\Delta/3$ the network period becomes smaller than the minigap, and
the situation resembles the one encountered when the voltage decreased below
$2\Delta$: we have to calculate the next correction to the DOS $\sim W^2$. In
the equation for the four-particle current,
\begin{equation}\label{I4>}
I_4 = \frac{8}{eR}\int_0^{2eV-\Delta} \frac{\rmd E} {\rho_{-1} + \rho_0 +
\rho_1 + \rho_2}, \qquad \frac{\Delta}{2} \leq eV < \frac{2\Delta}{3}
\end{equation}
(see figure \ref{DOS}(c)), the largest resistances are $\rho_0 \sim \rho_1
\sim W^{-3}$, thus $\rho_{-1} \sim \rho_{-2} \sim W^{-1}$ can be neglected.
The functions $N_{\pm 1}$ can be obtained from the SPT equations
\eref{DOSSPT} at $E=E_{\pm 1}$, in which small $N$ in the rhs has to be
neglected: $N_{\pm 1} = W| \Delta/\xi_{\pm 1}|^{a} N_{\rm S,\pm 2}$. To
evaluate the function $N$, we must perform next iteration step, by
substituting these values of $N_{\pm 1}$ into the SPT equation \eref{DOSSPT}
for $N$. As the result, we obtain
\begin{eqnarray}\nonumber
I_4 &= \frac{8 W^3}{eR} \int_0^{2eV-\Delta} \rmd
E\left|\frac{\Delta^3}{\xi\xi_{-1} \xi_1}\right|^{a} {\frac{|E_2
E_{-2}|}{\xi_2 \xi_{-2}}}.
\end{eqnarray}
While approaching the voltage value $eV = 2\Delta/3$, the current $I_4$
infinitely increases. In the vicinity of the threshold $eV = \Delta/2$ we
have
\begin{eqnarray} \label{I4thr}
I_4(\Delta/2) = \frac{2\pi \Delta W^3}{eR} \left( \frac{4}{3}
\right)^{a}  \approx 2W I_3(2\Delta/3).
\end{eqnarray}

From these considerations we conclude that the evaluation of $2n$- and
$(2n+1)$-particle currents requires DOS recurrences of $n$th order. As long
as the applied voltage decreases below $\Delta/n$, a new minigap opens in the
DOS (see discussion in section \ref{SecSPT}), and the recurrent procedure
should be repeated again.

\section{Multiparticle currents: large voltage}\label{Secpeaks}

It follows from the previous section that the multiparticle currents
calculated within the SPT approach have finite values in the vicinity of
their thresholds, but they diverge at the next gap subharmonics: the
two-particle current diverges at $eV=2\Delta$, the three-particle current
diverges at $eV=\Delta$, and so on. These divergences are caused by
singularity of the SPT correction to the tunnelling DOS at the point
$E=\Delta$ which enters the integration region.

It is easy to see that the two-particle current persists also above the gap
voltage, $eV>2\Delta$: when the node $n_0$ is inside the gap, $|E|<\Delta$,
the subgap circuit should consist of two resistors no matter how large the
applied voltage is. Furthermore, since the singular point $E=\Delta$ always
belongs to the integration region, the current will formally diverge at {\em
all voltages} $eV>2\Delta$; generally, the $n$-particle current ($n>1$) taken
in the SPT approximation diverges at all voltages above $eV=2\Delta/(n-1)$.
This catastrophe is known since the pioneering calculations of the
two-particle current within the tunnelling Hamiltonian model \cite{MPT} but
it can be eliminated by using the improved perturbation expansion of section
\ref{SecIPT}. This implies in fact that the currents have nonanalytical
dependencies on the tunnelling parameter $W$ at large voltages.

\begin{figure}[tb]
\begin{center}
\epsfxsize=8cm\epsffile{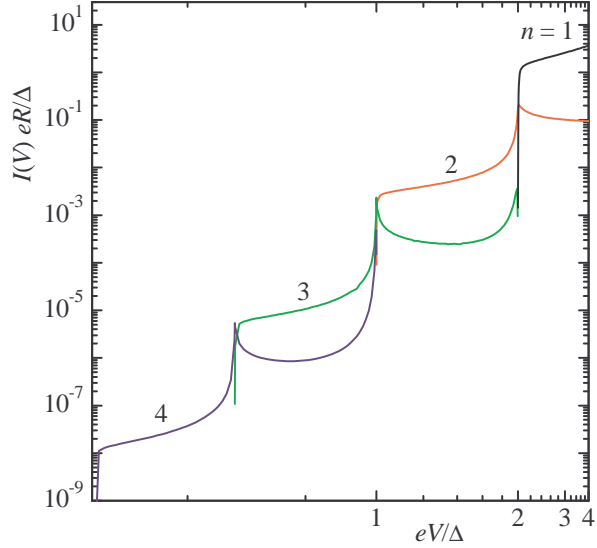} \end{center} \vspace{-5mm}
\caption{Multiparticle currents in planar junction numerically computed for
the tunnelling parameter $W=10^{-3}$.} \label{partial}
\end{figure}

We also note that the three-particle current disappears at large enough
voltage, in contrast to the two-particle current which persists at all
voltages above $\Delta/e$. This is obvious from the circuit geometry in
figure \ref{DOS}: as soon as $eV$ exceeds $2\Delta$, the network period
becomes larger than the energy gap, therefore the subgap circuit may involve
no more than two resistors. This is relevant for all $n$-particle currents
with $n>2$ which persist only within the voltage intervals $2\Delta/n < eV <
2\Delta/(n-2)$ and abruptly disappear at larger voltages (see figure
\ref{partial}).

\subsection{Two-particle current at $eV \geq 2\Delta$}

The two-particle current at $eV > 2\Delta$ is given by equation
\begin{equation} \label{I2}
I_2 = \frac{4}{eR}\int_0^{\Delta} \frac{\rmd E}{\rho_0 + \rho_1} =
\frac{4}{eR}\int_0^\Delta \rmd E \frac{N N_1 N_{-1}}{N_1 + N_{-1}}.
\end{equation}
To evaluate the integral, we use the IPT equations
\eref{IPT_2not}--\eref{IPT_2DOS} to calculate $N$, while the functions
$N_{\pm 1}$ are taken in the BCS form. Furthermore all the smooth functions
can be approximated with their values at the singular point $E=\Delta$,
\numparts\label{gh}
\begin{eqnarray}\label{g}
g(eV) = \case12\sum\nolimits_{k=\pm 1}N_{S}(\Delta+keV),
\\
N_1^{-1}+N_{-1}^{-1}=\sum\nolimits_{k=\pm 1}N_{S}^{-1}(\Delta+keV)\equiv
h^{-1}(eV).
\end{eqnarray}
\endnumparts
As it was expected, the two-particle current in this region is given by a
nonanalytical expression with respect to $W$, and it is larger than the
current value at smaller voltages $eV < \Delta$,
\begin{eqnarray}\label{I2large}
I_2 = \frac{2\Delta}{eR} C_1 p(eV) h(eV) \sim W^b ,
\\
C_1 = \int_0^{\infty} \rmd\epsilon \,\im z(\epsilon) = \cases{ 9\sqrt{3}/4,
& \textrm{1D junction,} \\
\sqrt{2}, & \textrm{planar junction.}} \nonumber
\end{eqnarray}
Here $p(eV)$ is given by (\ref{IPT_2not}) or \eref{p_planar} depending on the
junction geometry.

Important property of the tunnelling IVC is the excess current
$I_{\rm{exc}}$, i.e., voltage-independent deviation of the total current from
the ohmic IVC at large voltage $eV \gg \Delta$. The excess current is readily
evaluated by considering this limit in \eref{gh} and \eref{I2large},
\begin{equation}\label{IexcIPT}
I_{\rm{exc}} = \frac{\Delta}{eR} W^{b} \times\cases{
6.19, & \textrm{1D junction,} \\
2.83, & \textrm{planar junction.}}
\end{equation}

Our analysis is not complete yet because the current in \eref{I2large} grows
infinitely when the voltage approaches $2\Delta$. This divergence, caused by
the singularity of $N_{-1}$ in \eref{g}, can be eliminated by using a more
accurate approximation \eref{IPT_Notations}--\eref{DOS_IPT} for $N$.
Neglecting a nonsingular term $N_1$ in the denominator in \eref{I2} and
approximating $N_1$ with $N(3\Delta) = 3/2\sqrt{2}$, we obtain
\begin{eqnarray}
I_2(2\Delta) \approx \frac{4}{eR}\int_0^\Delta \rmd E \; N N_1 =
\frac{6q\Delta}{\sqrt{2}eR} \int_0^\infty \rmd\epsilon \re z(-\epsilon)
\nonumber
\\
\approx \frac{\Delta}{eR} W^{1/a}\times \cases{ 2.32, & \textrm{1D junction,} \\
2.50, & \textrm{planar junction.}}\label{deltaI_IPT1}
\end{eqnarray}
Thus we see that the two-particle current possesses a pronounced peak at
$eV=2\Delta$, which exceeds not only the current threshold value, but also
the large excess current, see figure \ref{partial}.

\subsection{Three-particle current at $ eV \geq \Delta$}

The three-particle current at $eV\geq \Delta $ is given by \eref{I3>}, in
which the integration interval is now $eV-\Delta <E < \Delta$. Using the
symmetry relation $N(E) = N(-E)$, we reduce the integration region to the
interval $eV/2 <E < \Delta$, containing only one dangerous point $E =
\Delta$,
\begin{equation}
I_3 =  \frac{6}{eR}\int_{eV/2}^{\Delta} \frac{\rmd E}{\rho_{-1} + \rho_0 +
\rho_1}.\label{I3}
\end{equation}
At this point, all terms in the denominator turn to zero being calculated
within the SPT approximation. To eliminate the divergence, we apply the IPT
approach, \eref{IPT_2DOS} and \eref{IPT_2not}, to calculate $N$, which then
acquires a finite value $\sim W^{-b}$ at the singular point. In the present
case the function $g$ defined in \eref{IPT_I2} has a complex-valued form
$g(eV) = |g| \rme^{\rmi\varphi}$ because the energy $E_{-1}$ in this equation
occurs inside the gap,
\begin{eqnarray}\label{g_complex}
|g| =\frac{\Delta^{3/2}}{\sqrt{eV \left[ (2\Delta)^2 -\! (eV)^2 \right]}},
\qquad \tan \varphi = \frac{eV - \Delta}{eV + \Delta}\sqrt{\frac{2\Delta +
eV}{2\Delta - eV}}.
\end{eqnarray}
Thus the scaling factor $p$ in \eref{p_Notations}, \eref{p_planar} and
\eref{IPT_2DOS} is to be defined through the modulus of $g$, while the phase
factor will remain in the equation for $z$ in the 1D geometry,
\begin{equation} \label{Eqzphi}
z^3 \exp (2\rmi\varphi) + \left(z\sqrt{\epsilon}-1 \right)^2 = 0,
\end{equation}
and in the expression for $z$ in planar geometry,
\begin{equation} \label{Eqz_planar}
z=[\epsilon-\rmi\exp(\rmi\varphi)]^{-1/2}.
\end{equation}
Therefore the normalized spectral function becomes dependent on
$eV$. The function $N_{-1}$ can then be evaluated from the SPT
approximation, keeping only large quantity $N$ in the rhs,
\begin{equation} \label{Nminus1}
N_{-1} = W |\Delta/\xi_{-1}|^{a} N.
\end{equation}
At the singular point, $N_{-1} \sim W^{1-b}$; the other relevant functions
$N_1$ and $N_{-2}$ can be taken in the BCS form. Retaining only the largest
resistance $\rho_{-1}\sim W^{b-1}$ in \eref{I3}, we get
\begin{eqnarray}\nonumber
I_3 = \frac{6}{eR} \int_{eV/2}^\Delta \rmd E \, N_{-1} N_{-2} =
\frac{6Wp\Delta}{eR }N_{-2} \left|\frac{\Delta}{\xi_{-1}}\right|^{a} C_2 \sim
W^{1+b},
\\
C_2(eV) = \int_0^{\infty} \rmd\epsilon \,\im z(\epsilon, eV) \label{I3a}
\end{eqnarray}
(for the planar model, $C_2(eV)=2\cos(\varphi/2 +\pi/4)$). Here the functions
$N_{-2}$ and $\xi_{-1}$ should be taken at the point $E=\Delta$. This
expression diverges at $eV = \Delta$ and $eV = 2\Delta$. In fact, the current
has a finite peak value at $eV = \Delta$; another peak appears slightly below
$eV = 2\Delta$, because the current turns to zero at $eV = 2\Delta$, as
follows from \eref{I3} (see figure \ref{partial}).

At $eV = \Delta$, the function $N_{-2}$ is also large at the point $E=\Delta$
and should be kept in \eref{Nminus1} together with $N$, i.e., $N_{-1} = W (N
+ N_{-2})$; both these functions are to be evaluated using the IPT scheme
\eref{IPT_I2}--\eref{IPT_2DOS}. This leads to expressions $N = (2p)^{-1} \im
z_+(\epsilon)$ and $N_{-2} = (2p)^{-1} \re z_-(\epsilon)$, where the
functions $z_\pm$ are given by the solutions of algebraic equations
\begin{eqnarray}\label{Eqzpm}
z^3_+ + (z_+ \sqrt{\epsilon}-1)^2=0,\qquad \rmi{z}^3_- + ({z}_-
\sqrt{\epsilon}-1)^2=0
\end{eqnarray}
in the case of 1D junction, or by explicit expressions
\begin{eqnarray}\label{Eqzpm_planar}
z_\pm = (\epsilon\mp \rmi)^{-1/2}
\end{eqnarray}
for a planar junction; the scaling parameter $p$ is $(W^2/6)^{1/3}$ and
$(W/\sqrt 3)^{1/2}$, respectively. Then the largest resistances are $\rho_{0}
\sim \rho_{-1} \sim W^{2b-1}$, and we obtain
\begin{eqnarray}
I_3(\Delta)= \frac{6W}{eR}\int_{\Delta/2}^\Delta {\rmd E\, N\, N_{-2}}=
\frac{3W\Delta }{eR}\int_0^\infty \rmd\epsilon\,\im z_+ \re {z}_- \nonumber
\\
= \frac{ W \Delta}{eR} \times \cases{ 3.16, & \textrm{1D junction;} \\
3\pi/4, & \textrm{planar junction.}}\label{I3_6}
\end{eqnarray}

Analysis of the current behaviour near $eV=2\Delta$, which we do not present
here, gives the following estimate for the current peak value: $I_{3\rm{max}}
\sim W^{4/5}\Delta/eR$ for the 1D junction and $I_{3\rm{max}} \sim
W\Delta/eR$ for the planar junction.

\subsection{Four-particle current at $eV \geq 2\Delta/3$}

The four-particle current at $eV \geq 2\Delta/3$ is given by \eref{I4>},
where the upper integration limit is replaced by the value $\Delta-eV$,
representing a singular point of the integrand. The functions $N$ and
$N_{-1}$ are calculated from the SPT equations \eref{DOSSPT}, neglecting
their values in the rhs,
\begin{eqnarray}\label{N4}
N = W| \Delta/\xi|^{a} N_1, \qquad N_{-1} = W|\Delta/\xi_{-1}|^{a} N_{-2}.
\end{eqnarray}

Within the voltage interval $2\Delta/3 < eV < \Delta$ we can apply the BCS
approximation for the functions $N_{\pm 2}$. However, the function $N_1$ must
be calculated by means of the IPT \eref{IPT_I2}--\eref{IPT_2DOS}, as soon as
the energy $E_1$ hits the singular point. The resulting equations for $N_1$
coincide with \eref{g_complex}--\eref{Eqz_planar} for $N$ in the previous
section. Thus, at the singular point, $N_1 \sim W^{-b}$, $N_0 \sim W^{1-b}$,
$N_{-1} \sim W$, and $N_{\pm 2} \sim 1$. Therefore the resistance $\rho_0
\sim W^{b-2}$ dominates, which gives
\begin{eqnarray}
I_4 = \frac{8 \Delta W^2}{eR} C_2(eV)p(eV) \left|\frac{\Delta^2}{\xi
\xi_{-1}} \right|^{a} N_{-2} \sim W^{2+b},
\end{eqnarray}
where the functions $\xi$, $\xi_{-1}$, and $N_{-2}$ are taken at
$E=\Delta-eV$. This expression diverges at the points $eV = 2\Delta/3$ and
$eV = \Delta$, where the IPT should be applied. Since the SPT approximation
for $N_{-2}$ diverges at $eV = 2\Delta/3$, we evaluate both the functions
$N_1$ and $N_{-2}$ in the rhs of \eref{N4} using the IPT equations
\eref{IPT_I2}--\eref{IPT_2DOS},
\begin{eqnarray}
N_1(E) = (2p)^{-1} \im z_+(\epsilon),\qquad N_{-2}(E) = (2p)^{-1} \re
{z}_-(\epsilon).
\end{eqnarray}
The functions $z_\pm$ obey the equations \eref{Eqzpm} or \eref{Eqzpm_planar}
modified by the phase factors similar to \eref{Eqzphi} and \eref{Eqz_planar}.
In this case, $\tan \varphi =- \sqrt{2}/5$, and the scaling factor is
$p=\case34(W^2/2)^{1/3}$ for 1D junctions and $p=(3W\sqrt{3}/8)^{1/2}$ for
planar junctions. The corresponding estimates for the DOS functions are $N_1
\sim N_{-2} \sim W^{-b}$, $N \sim N_{-1} \sim W^{1-b}$, while $N_2 \sim 1$
can still be taken in the BCS approximation. Therefore the largest resistance
is $\rho_0 \sim W^{2b-2}$, and we get
\begin{eqnarray}
I_4(2\Delta/3) = \frac{4 \Delta W^2}{eR} \left(\frac{9}{8}\right)^a
\int_0^\infty \rmd\epsilon \im z_+ \re {z}_- \nonumber
\\
=\frac{\Delta W^2}{eR}\times \cases{ 6.63, & \textrm{1D junction;} \\
5.26, & \textrm{planar junction.}}
\end{eqnarray}

Analysis of the current peak near $eV = \Delta$ gives the estimate
$I_{4\rm{max}} \sim W \Delta/ eR$.

\section{Effect of first harmonics}\label{Secharmonics}

In this Section we evaluate the effect of higher harmonics on the dc subgap
current. We restrict our calculation to first order in the tunnelling
parameter $W$ thus taking into account only two first harmonics $m = \pm1$.
The main conclusion of our calculation will be that including harmonics
produces just insignificant quantitative changes, while major qualitative
properties of the SGS, positions and scaling of the current steps will not
change. We start with a general equation \eref{I0} for the dc current, and
evaluate an additional contribution due to the first harmonics of the
boundary values of the Keldysh and Green's functions,
\begin{eqnarray}
\delta I =\int_{-\infty}^{\infty} \frac{\rmd E}{32 eR}
\Tr\sum\nolimits_{m=\pm1} m\,\sigma_z \, \bigl[\hat{h}(E, 0) \hat{G}^{\rm
K}(E, m) \nonumber
\\
+ \hat{h}(E, - m)\hat{G}^{\rm K}(E, 0)\bigr] \label{delta_I2}
\\
= \rmi\int_{-\infty}^{\infty}  \frac{\rmd E}{4eR}\sum\nolimits_{m=\pm1} m
\,\left [V_{y}(E, m)\,\im v + V \,\im v_y(E, m)\right].\nonumber
\end{eqnarray}
Here and below we use the subscripts $(x,y,z)$ to indicate the matrix
components of the first harmonics of the Green's and Keldysh functions, while
the zero harmonics will be used as before without such subscripts. Thus $v$
[$V$] in \eref{delta_I2} indicates $y$-component of $\hat g(E,0)$ [$\hat
G^{\rm K}(E,0)$], and $v_{y}(E,\pm1)$ [$V_{y}(E,\pm1)$] indicates the
$y$-component of $\hat g(E,\pm1)$ [$\hat G^{\rm K}(E,\pm1)$].

\subsection{Perturbation theory for the Green's functions}

To evaluate the current in \eref{delta_I2}, one needs to find the
$y$-component in the Green's function expansion over the Pauli matrices,
$\hat{g}(z,E, m) = \sigma_z u_z + \rmi\sigma_y v_y + \sigma_x v_x + w$. In
what follows, we will focus on the case of 1D junction. Considering the
Usadel equation \eref{Usadel} to first order in $W$, we arrive at the
equation for the functions $v_y(z,E, m)$ and $u_z(z,E, m)$,
\begin{equation}
m eV u_z = 2\rmi \Delta\left(u_m u_z^{\prime\prime} - v_m
v_y^{\prime\prime}\right).\label{equ_z}
\end{equation}
In this and following derivations we neglect small deviation of the functions
$u$ and $v$ from the BCS form \eref{xi}. We use the following convention: the
zero harmonics with shifted energy arguments $E+meV$ will be denoted with the
subscript $m$, as before, e.g., $u_m \equiv u(E+meV,0)$, while the first
harmonics will be denoted with an argument, e.g., $v_y(E,\pm1)$; the absence
of the explicit argument would mean the relevance to the both harmonics
$m=\pm 1$.

The function $u_z$ can be excluded from \eref{equ_z} by virtue of the
normalization condition $(u_1 + u_{-1})u_z =(v_1 + v_{-1})v_y$ following from
\eref{Usadel}. The boundary condition for \eref{equ_z} results from
\eref{BoundaryG} and determines the boundary value of first derivative,
\begin{equation}
v^\prime_{y}|_{z=-0} = ({W}/{2}) v (1 + u_1 u_{-1} + v_1
v_{-1}).\label{v_yb}
\end{equation}
Solution of \eref{equ_z}, $v_y(z) = v_{y} \exp(k_1 z)$ at $z<0$, leads to the
following relation at the boundary,
\begin{eqnarray}\label{deltag}
v_{y}(E,m)=k_1^{-1}v^\prime_{y}(E, m), \qquad k_1^2 = (\xi_1 +
\xi_{-1})/2\rmi\Delta.
\end{eqnarray}
Equations \eref{v_yb} and \eref{deltag} allow us to express the first
harmonics $v_{y}(E,m)$ through known zero harmonics $u$ and $v$, and to
establish the relation $v_{y}(E, 1) = v_{y}(E, -1)$ which implies that the
second term in \eref{delta_I2} turns to zero.

\subsection{Perturbation theory for Keldysh function}

To calculate the harmonics of the Keldysh function, it is convenient to
separate the contributions of the harmonics of the Green's function and the
distribution function,
\begin{equation}\label{Keld_expand}
\eqalign{ \hat{G}^{\rm K}(E,m) = \delta\hat{G}^{\rm K}(E,m) +
\hat{\mathcal{G}}^{\rm K}(E,m),
\\
\delta\hat{G}^{\rm K}(E,m) =\hat{g}^{\rm R}(E,m) f_{-m} - f_m \hat{g}^{\rm
A}(E,m),
\\
\hat{\mathcal{G}}^{\rm K}(E,m) = \hat{g}^{\rm R}_m {f}(E,m) - {f}(E,m)
\hat{g}^{\rm A}_{-m}.}
\end{equation}
In \eref{Keld_expand} we used the rule \eref{harm} for the convolutions valid
for the first harmonics, $(A \circ B)(E, m) = A(E, m) B_{-m} + A_m B(E, m)$,
which allows us to write the equation for the function
$\hat{\mathcal{G}}^{\rm K}$ following from \eref{KineticG1} and
\eref{BoundaryGK} in a symbolic form
\begin{equation}
\bigl[ \hat{H}, \circ \hat{\mathcal{G}}^{\rm K}\bigr] = 2\rmi \Delta
\partial_x \bigl( \hat{g}^{\rm R} \circ \partial_x \hat{\mathcal{G}}^{\rm K} +
\hat{\mathcal{G}}^{\rm K} \circ \partial_x \hat{g}^{\rm A}
\bigr),\label{G1eq}
\end{equation}
where a small gradient of the distribution function, $\partial_x f \sim
L^{-1}$, has been neglected. The boundary condition to \eref{G1eq} reads
\begin{equation}
\hat{g}^{\rm R} \circ \partial_x \hat{\mathcal{G}}^{\rm K} = W\bigl(
\hat{g}^{\rm R} \circ \hat{\mathcal{F}} - \hat{\mathcal{F}} \circ
\hat{g}^{\rm A} \bigr),\label{bound_2K}
\end{equation}
where $\hat{\mathcal{F}} = \overline{\hat{g}^{\rm R}} \circ ( \overline{f} -
f ) - ( \overline{f} - f ) \circ \overline{\hat{g}^{\rm A}}$. According to
\eref{delta_I2} and \eref{Keld_expand}, we are interested in the
$y$-component in the expansion $\hat{\mathcal{G}}^{\rm K} = \sigma_z
\mathcal{U}_z + \rmi\sigma_y \mathcal{V}_y + \sigma_x \mathcal{V}_x +
\mathcal{W}$. Equation for this component, $m eV \mathcal{U}_z = 2\rmi\Delta
\left( u_m \mathcal{U}_z''
 - v_m \mathcal{V}_y'' \right)$, follows from \eref{G1eq} and looks
similar to \eref{equ_z}. The normalization condition \eref{G_Knorm1} gives
the relation $(v_m - v^*_{-m})\mathcal{U}_z=(u_m - u^*_{-m})\mathcal{V}_y$
which allows us to obtain a closed differential equation for $\mathcal{V}_y$.
Solution $\mathcal{V}_y(z) = \mathcal{V}_{y} \exp(q_1 z)$ of this equation at
$z<0$ gives the following relation at the boundary,
\begin{equation}\label{mathcalVy}
\mathcal{V}_{y }(E,m)= q_1^{-1}\mathcal{V}^\prime_{y}(E,m), \qquad q_1^2 =
(\xi_1 - \xi_{-1}^*)/2\rmi\Delta,
\end{equation}
where the quantity $\mathcal{V}_y^\prime$ is to be determined from the
boundary condition \eref{bound_2K},
\begin{eqnarray}\label{mathcalV'}
\mathcal{V}_{y}^\prime|_{z=-0} &= W\left( 1 - u_m u_{-m}^* - v_m v^*_{-m}
\right)\Phi(E,m),
\\
\Phi(E,m) &= M \Bigl( f - \case12\sum_{k = \pm 1} f_k \Bigr) +
 \case\rmi2 M_{\rm S}\sgn(m)  \sum_{k = \pm 1} k f_k,\nonumber
\end{eqnarray}
where $M_{\rm S} = \im v$.  At $T=0$ we obtain $\Phi(E,m) =
\Theta(eV-|E|)[M\sgn(E) + \rmi M_{\rm S}\sgn(m)]$.

\begin{figure}[tb]
\begin{center}
\epsfxsize=8cm\epsffile{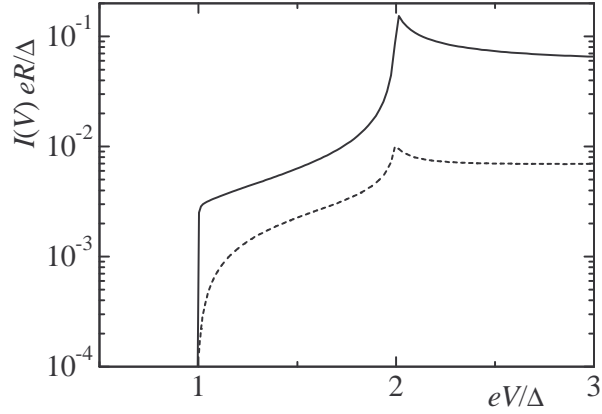} \end{center} \vspace{-5mm}
\caption{Current of first harmonics (dashed line) compared to the
two-particle current (solid line), for the tunnelling parameter $W=10^{-3}$.}
\label{harmonics}
\end{figure}

\subsection{Current of first harmonics}

The function $V_{y}(E, m)$, which determines the current in \eref{delta_I2},
is now expressed, according to \eref{Keld_expand}, through the quantities
calculated in \eref{v_yb}--\eref{deltag} and
\eref{mathcalVy}--\eref{mathcalV'}, $V_{y}(E, m) = v_{y} f_{-m} + v_{y }^*
f_m + \mathcal{V}_{y}$. Collecting all these equations and substituting them
into \eref{delta_I2}, we get
\begin{eqnarray}\nonumber
\delta I = \frac{W}{4eR} \int_{-\infty}^{\infty} \rmd E\, \im v \bigl\{ (f_1
- f_{-1})\im\bigl( v k_1^{-1}\cosh^2\chi\bigr) \nonumber
\\
+ \im\bigl[ v q_1^{-1}(f-f_{-1})+ v^* q_1^{-1}(f-f_1)}\sinh^2 {\widetilde\chi
\bigr]\bigr\},\nonumber
\end{eqnarray}
where $\chi = \case12(\theta_1+\theta_{-1})$ and $\widetilde\chi =
\case12(\theta_1+\theta_{-1}^*)$.  At $T=0$, this equation simplifies:
\begin{eqnarray}
\delta I = \frac{2W}{eR} \int_0^{eV} \rmd E\; \im v \im\bigl[v\bigl( k_1^{-1}
\cosh^2\chi +q_1^{-1} \sinh^2\widetilde\chi\bigr)\bigr]. \label{dI}
\end{eqnarray}

Similar considerations in the case of a planar junction result in
replacements $q_1 \to q_1^2$ and $k_1 \to k_1^2$ in the rhs of \eref{dI}.
Noting that at $eV < \Delta$ the energy $E_{-1}$ in \eref{dI} appears in the
subgap region, where $\theta^*_{-1} = \theta_{-1}+\rmi\pi$ and $\xi_{-1}^*
=-\xi_{-1}$, we find that $\delta I$ turns to zero at $eV < \Delta$, similar
to $I_2$. Numerical calculations show that the contribution of the first
harmonics to the net dc current does not exceed $30\%$ (see figure
\ref{harmonics}). From this we conclude that the adopted quasi-static
approximation, where the nonzero harmonics are neglected, gives a relatively
good approximation to a complete solution.

\section{Discussion}\label{Diss}

\begin{figure}[tb]
\begin{center}
\epsfxsize=8cm\epsffile{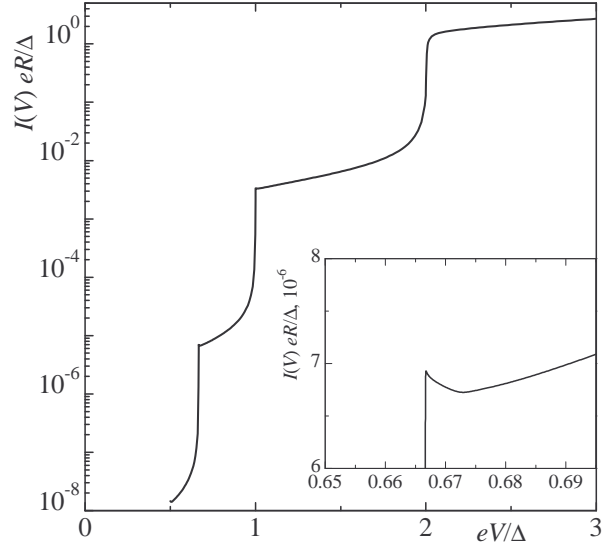} \end{center} \vspace{-5mm}
\caption{$I$-$V$ characteristic with tunnelling SGS for planar junction
computed numerically for the tunnelling parameter $W=10^{-3}$. The inset
shows a spike of the IVC near the threshold $eV = 2\Delta/3$ of the
three-particle current.} \label{iv1}
\end{figure}

\Table{\label{table1} Threshold and peak values of the normalized
multiparticle currents $I_n eR/\Delta$, $n=1 \div 4$. The left and right
sub-columns correspond to the 1D and planar models of the tunnel junction,
respectively.}
\br
$n$ & $eV= 2\Delta/n$ & $eV= 2\Delta/(n-1)$ & $eV= 2\Delta/(n-2)$ \\
\mr
1 & ${\pi}/{2}$ & --- & --- \\
2 & $\pi W$ & $2.32W^{2/5} \quad 2.50W^{1/3}$ & $6.19W^{1/3} \quad
2.83{W}^{1/2}$ \\
3 & $6.33W^2 \quad 6.71W^2$ & $3.16W \quad\;\;\;\;\;2.36W$ & $0.44W^{4/5}
\quad 3.57 W$ \\
4 & $12.9 W^3  \quad 14.9W^3 $ & $6.63W^2 \quad\;\;\;5.26 W^2$ & $0.57W
\quad\;\;\;\;\;\,0.48W$ \\
\br
\endTable

Our analysis of the high-order multiparticle currents shows that they exhibit
a similar pattern of the voltage dependence (see figure \ref{partial}): an
$n$-particle current appears above the threshold $eV= 2\Delta/n$, having
roughly the value $I_{n} \sim (2W)^{n-1} I_1$, then increases and shows
dramatic peak while approaching $eV= 2\Delta/(n-1)$; then it remains
anomalously large within the voltage interval $2\Delta/(n-1) < eV<
2\Delta/(n-2)$ and eventually disappears at $eV = 2\Delta/(n-2)$, showing
another strong peak at slightly smaller voltage. For convenience, all the
threshold and peak values of the first four currents are brought together in
the table \ref{table1}.

The net tunnel current consists of the sum of the $n$-particle currents, and
therefore exhibits a pronounced step-like structure on the IVC with steps at
the gap subharmonics $eV=2\Delta/n$, as shown in figure \ref{iv1} obtained by
numerical calculation at $T=0$. The peaks of the multiparticle currents with
numbers $n+1$ and $n+2$ produce small spikes at the $n$th threshold with
$n>1$; the example of such spike is presented in the inset of figure
\ref{iv1}. The numerical procedure involves solving the set of recurrences
\eref{Eqtheta1D} or \eref{EqthetaPl} for the functions $\theta_k $ which
correspond to the subgap energies, $|E_k| \leq \Delta$ ($-N_- < k < N_+$);
the nonsingular terms in these equations are replaced with the BCS values.
For the voltage values equal to the gap subharmonics one more equation is to
be added as explained in the previous sections. We note that the results for
the 1D and planar geometries differ insignificantly for equal values of the
tunnelling parameter \eref{W} and \eref{Wtilde}; in logarithmic scale, the
difference can be detected only in the immediate vicinity of the peaks.

Such a picture is quite similar to the tunnelling SGS in quantum point
contacts \cite{Bratus95,Cuevas96,Averin95,Post94,Jan2000}, and the resulting
IVC is found to be very close to the result for a point contact with the
effective transparency $D_{\rm{eff}} = 4W$. Thus, according to the
definitions \eref{W} and \eref{Wtilde} of the tunnelling parameter $W$, the
enhancement factor $D_{\rm{eff}}/D$ for the SGS scaling is equal to
$3\xi_0/\ell$ in the 1D junction and $3\xi_0^2/d\ell$ in the planar junction.
In particular, for planar Al junctions with $\ell \sim d = 50$ nm and $\xi_0
= 300$ nm, the enhancement factor may approach the value 100.

The fact that the SGS in planar junctions is sensitive to the properties of
the junction electrodes has important implications for characterization of
the junction tunnelling layer. Indeed, the thickness of this layer in
realistic junctions is inhomogeneous: there are spots with enhanced
transparency, which mostly contribute to the tunnel current. If the linear
sizes of such spots are large compared to the electron mean free path in the
electrodes (in practice, the thickness of thin-film electrodes), the junction
can be considered as a quasi-planar one, and the SGS should be enhanced
according to our theory and depend on the electrode thickness. However, if
such spots are small compared to the electron mean free path, the tunnel
current rapidly spreads out in immediate vicinity of the spot without being
affected by the impurity scattering, as it is in the ballistic constrictions;
in this case, there must be no dependence of the SGS on the electrode
properties in accord with the mesoscopic theory prediction
\cite{Bratus95,Cuevas96,Averin95}.

\ack{The support from the SSF-OXIDE Consortium, the Swedish Research Council
(VR), and the Royal Academy of Science (KVA) is gratefully acknowledged.}

\end{document}